\documentclass[%
 reprint,
 amsmath,amssymb,
 aps,
]{revtex4-2}

\usepackage{graphicx}% Include figure files
\usepackage{dcolumn}% Align table columns on decimal point
\usepackage{bm}% bold math
\usepackage{xcolor}
\begin{document}

\preprint{APS/123-QED}

\title{Collective rovibronic dynamics of a diatomic gas
coupled by cavity}% Force line breaks with \\
%\thanks{A footnote to the article title}%

\author{Niclas Krupp}
 \email{niclas.krupp@pci.uni-heidelberg.de}
 \affiliation{%
 Theoretische Chemie, Physikalisch-Chemisches Institut, Universität Heidelberg, INF 229, 69120 Heidelberg, Germany 
}%
 %\altaffiliation[Also at ]{Physics Department, XYZ zu University.}%Lines break automatically or can be forced with \\
\author{Oriol Vendrell}%
 \email{oriol.vendrell@uni-heidelberg.de}
\affiliation{%
 Theoretische Chemie, Physikalisch-Chemisches Institut, Universität Heidelberg, INF 229, 69120 Heidelberg, Germany 
}%

\date{\today}% It is always \today, today,
             %  but any date may be explicitly specified

\begin{abstract}
We consider an ensemble of homonuclear diatomic molecules coupled to the two polarization directions of a Fabry-Pérot cavity via fully quantum simulations. Accompanied by analytical results, we identify a coupling mechanism mediated simultaneously by the two perpendicular polarizations, and inducing polaritonic relaxation towards molecular rotations. This mechanism is related to the concept of light-induced conical intersections (LICI). However, unlike LICIs, these non-adiabatic pathways are of collective nature, since they depend on the \emph{relative} intermolecular orientation of all electronic transition dipoles in the polarization plane. Notably, this rotational mechanism directly couples the bright upper and lower polaritonic states, and it stays in direct competition with the collective relaxation towards dark-states. Our simulations indicate that the molecular rotational dynamics in gas-phase cavity-coupled systems can serve as a novel probe for non-radiative polaritonic decay towards the dark-states manifold.
\end{abstract}

%\keywords{Suggested keywords}%Use showkeys class option if keyword
                              %display desired
\maketitle

%\tableofcontents
Strong coupling between matter excitations and quantized electromagnetic modes leads to the formation of hybrid light-matter states, polaritons \cite{weisbuch_observation_1992,torma_strong_2015}. Due to their hybrid nature, polaritons possess new properties differing from the bare material system \cite{ebbesen_hybrid_2016}, which has stimulated research across the natural sciences \cite{basov_polariton_2020}, including ultracold physics \cite{kasprzak_boseeinstein_2006, sun_bose-einstein_2017}, material sciences \cite{hagenmuller_cavity-enhanced_2017, balasubrahmaniyam_enhanced_2023}, molecular physics \cite{cho_optical_2022, xiang_intermolecular_2020} and chemistry \cite{feist_polaritonic_2018}.
% Although coupling molecular excitations to the quantized radiation field inside optical cavities has resulted in major experimental breakthroughs such as cavity-suppressed \cite{ahn_modification_2023} or accelerated \cite{hutchison_modifying_2012} chemical reactions and cavity-mediated energy transfer \cite{coles_polariton-mediated_2014}, a comprehensive theory that can adequately describe or even predict experimental results is still elusive \cite{fregoni_theoretical_2022}. 
%
However, important questions remain elusive in the field of polaritonic chemistry,
including the role of collective and local effects, molecular decoherence pathways, and cavity losses \cite{fregoni_theoretical_2022,sidler_perspective_2022}.
This significant difficulty originates from the contrast between theoretically simulated
scenarios, which navigate a compromise between small isolated systems
and model approximations in fundamental theories,
and the majority of experiments conducted in solution and condensed phases.

The recent advent of an experimental platform for the formation of gas-phase molecular polaritons in Fabry-Pérot (FP) cavities
enables studying molecules under strong coupling conditions with a high level of control, and over wide temperature ranges \cite{wright_rovibrational_2023, wright_versatile_2023}.
These platforms can provide much sought-after experimental insight into fundamental mechanisms through which molecules and quantized electromagnetic modes interact, and the basis for comparisons to accurate quantum-dynamics simulations of molecular ensembles.
In light of these developments, we study the full-dimensional quantum dynamics of prototypical gas-phase diatomic molecules inside a FP cavity by efficient numerical propagation of the time-dependent Schrödinger equation for the cavity-ensemble system with the ML-MCTDH method \cite{vendrell_multilayer_2011, vendrell_coherent_2018}. Our simulations highlight the competition between
several nonradiative relaxation pathways involving rotational, vibrational, and electronic degrees of freedom of
the ensemble. Most importantly, we characterize a new rotational collective coupling mechanism associated with the
relative orientation of the transition dipole moments in the ensemble. This channel couples the upper
and lower polaritons via conical intersections, and it stays in direct competition with the decay
to the dark-states manifold via vibrational collective conical intersections (CCI)~\cite{vendrell_collective_2018}.
The experimental verification of these competing mechanisms would constitute a leap forward in our microscopic
understanding of polaritonic photochemistry.

So far, theoretical studies have mainly focused on the role of the angle between the molecular transition dipole moment (TDM) vector and the cavity polarization direction of a one-dimensional FP cavity model \cite{fan_quantum_2023, triana_polar_2021, szidarovszky_conical_2018, triana_revealing_2019, sidler_chemistry_2020, flick_cavity-correlated_2018}. This rotational DOF has been shown to give rise to light-induced conical intersections (LICIs) in  molecules which are coupled to quantized electromagnetic modes \cite{szidarovszky_conical_2018, csehi_competition_2022}. Originally, LICIs were proposed for diatomic molecules exposed to strong classical laser fields where degeneracies between dressed potential energy surfaces are lifted along the polarization angle (with respect to the molecular axis), and the interatomic separation \cite{sindelka_strong_2011, szidarovszky_direct_2018}. Their rich nonadiabatic effects have been studied theoretically and experimentally, including topological phase effects \cite{halasz_light-induced_2012}, intensity borrowing and highly mixed rovibrational levels in absorption spectra \cite{szidarovszky_conical_2018, fabri_striking_2020} as well as  quantum interference in laser-induced photodissociation \cite{natan_observation_2016}.  

\begin{figure}[h]
\includegraphics[width=0.4\textwidth]{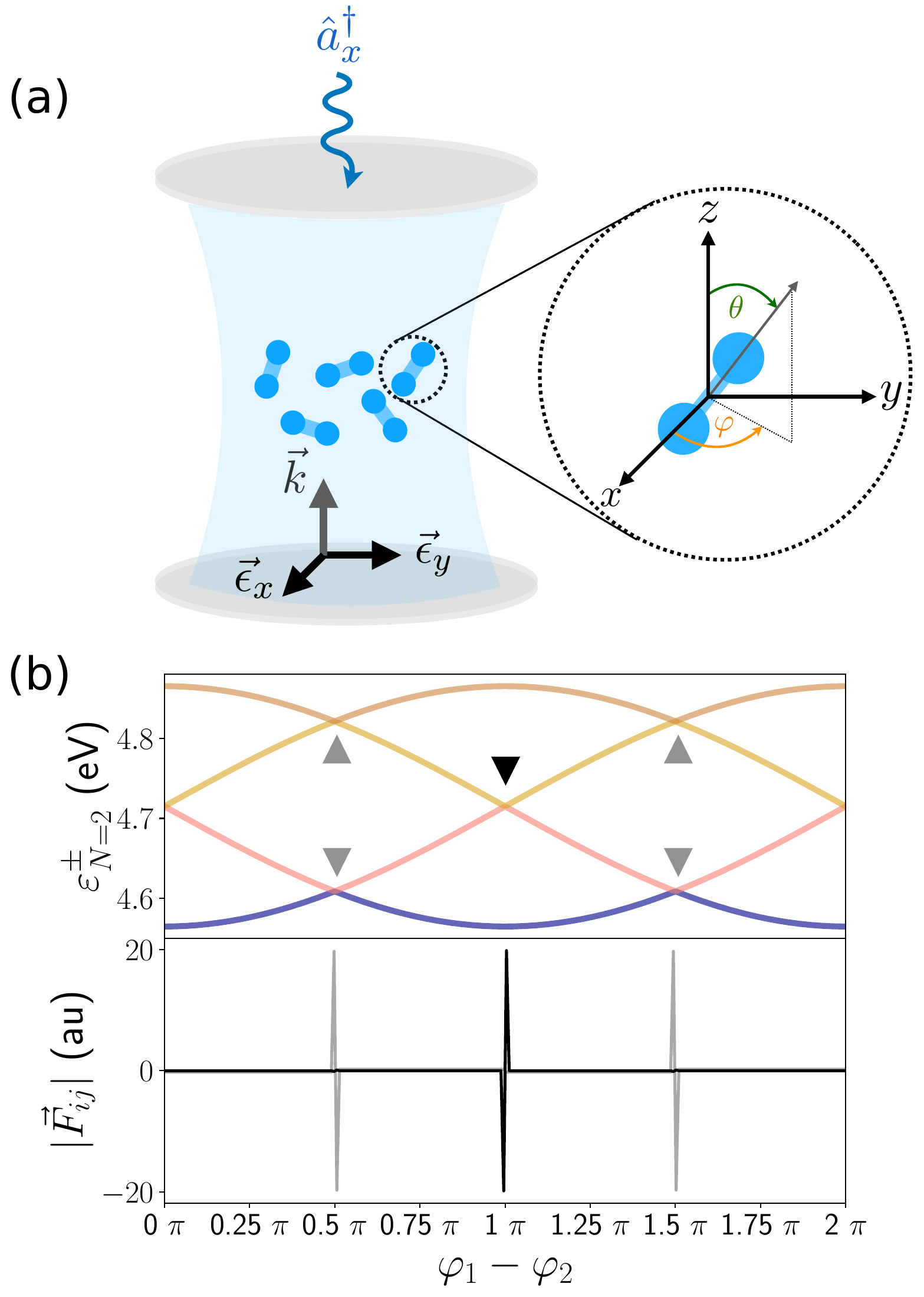}
\caption{(a) Schematic of a gas of homonuclear diatomic molecules inside a FP cavity with polarization directions $\vec\epsilon_x$ and $\vec\epsilon_y$. Space-fixed axes $x,y,z$ are parallel to $\vec\epsilon_x, \vec\epsilon_y, \vec k$. The orientation of molecular dipoles with respect to the space-fixed frame is parameterized by spherical coordinates $\theta$ and $\varphi$. The molecular rotational response after one-photon excitation of the $x$-polarized cavity mode is investigated. (b) Polaritonic potential energy curves for two molecules, $\varepsilon^{\pm}_{N=2}$, along relative angle $\varphi_1-\varphi_2$ (upper panel). Nonadiabatic couplings $\vec F_{ij}$ (lower panel) among the two LP or UP states (grey), and between LP and UP (black) diverge at respective degeneracies of pPECs (grey and black triangles). Parameters $\omega_C=\Delta^{(1)}=\Delta^{(2)}=4.71$\,eV and $\lambda=0.1/\sqrt{2}$\, eV were used.\label{fig:figure_theo} }
\end{figure}

A single angle is sufficient to characterize the orientation of a diatom with respect to a linearly-polarized laser field. In a realistic FP cavity, on the other hand, doubly degenerate cavity modes with orthogonal polarization directions are intrinsically present \cite{craig_molecular_1998}. Hence, two rotational DOFs must be included in order to account for all possible molecular orientations with respect to the $x$- and $y$-polarized cavity modes: the out-of-plane rotation along the polar angle $\theta_i$, and the in-plane rotation along the azimuthal angle $\varphi_i$ (where ``plane'' refers to the cavity polarization plane), as depicted in Fig.~\ref{fig:figure_theo}~(a).
While $\theta_i$ modulates the overall individual coupling strength of each molecule to the cavity,
the in-plane rotation $\varphi_i$ tunes the mixing between $x$-polarized and $y$-polarized cavity-mode contributions
to the molecule-cavity interaction.

 Crucially, the in-plane rotations $\varphi_i$ determine the collective dynamics of the
 ensemble in a way that the out-of-plane angles $\theta_i$ do not.
 We examine and explain this hitherto unknown collective mechanism on the basis of analytical considerations
 and numerical simulations.
 This leads to an extension of the notion of LICI to azimuthal angles,
 where the ensemble dynamics is determined by collective in-plane rotations of the transition dipole moments
 inside the FP cavity.
 In particular, conical intersections (CIs) among polaritons, whose branching
 spaces are spanned by molecular vibrations and rotations (both in- and out-of-plane plane),
 lead to ultrafast and nonresonant electronic-energy transfer to rotations.
 %which critically depends on population of bright polaritonic states.
 We note that the resonant scenario, in which a microcavity is directly coupled to a rotational transition has been studied recently on the single molecule level \cite{fan_quantum_2023}.
 
We consider an ensemble of $N$ non-interacting, homo-nuclear diatomic molecules
strongly coupled to two quantized electromagnetic modes with polarization directions $\vec\epsilon_x$ and $\vec\epsilon_y$ inside a FP cavity. The cavity-ensemble Hamiltonian for this systems reads \cite{vendrell_coherent_2018, fischer_cavity-induced_2022}
\begin{eqnarray}
    \hat H = \sum_{i=1}^N \hat H^{(i)}_{\mathrm{mol}} + \hat H_{\mathrm{cav}}, \label{eq:Ham-tot}
\end{eqnarray}
where $\hat H_{\mathrm{cav}}$ describes the cavity and cavity-ensemble interaction,
\begin{eqnarray}
    \hat H_{\mathrm{cav}} &=& \hbar\omega_c\left(1 +\hat a_x^{\dagger}\hat a_x  +\hat a_y^{\dagger}\hat a_y \right) \nonumber\\ && + g\sum_{i=1}^N  \hat{\vec\mu}^{(i)}\vec{\epsilon}_x \left(\hat a_x^{\dagger}+\hat a_x\right)  +\hat{\vec\mu}^{(i)}\vec{\epsilon}_y \left(\hat a_y^{\dagger}+\hat a_y\right),
\end{eqnarray}
with cavity frequency $\omega_c$, the coupling strength $g$ between the molecules and cavity modes, and the total dipole operator of the molecular ensemble $\sum_{i=1}^N \hat{\vec\mu}^{(i)}$. The quadratic dipole self energy term is neglected as it is only relevant for very strong coupling strengths which are not considered here \cite{flick_atoms_2017}. The validity of this approximation is examined and confirmed in the Supporting Material \cite{supp}.

\begin{figure*}
    \centering
    \includegraphics[width=\textwidth]{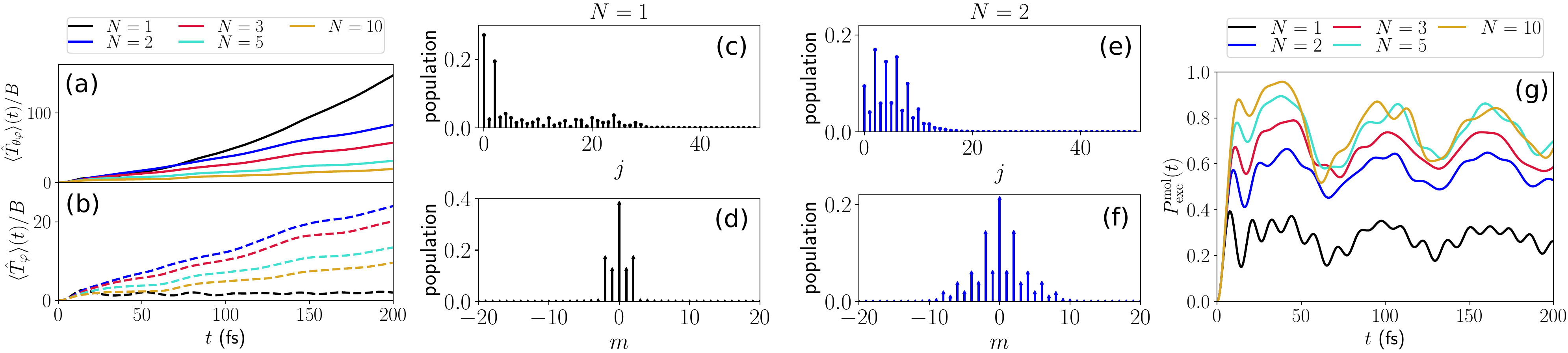}
    \caption{Molecular rotational dynamics after excitation of $x$-polarized cavity mode for various ensemble sizes. (a)-(b) Total kinetic rotational energy $\langle \hat T_{\theta\varphi}\rangle(t)$ and in-plane rotational energy  $\langle \hat T_{\varphi}\rangle(t)$. Energies are given in units of the rotational constant $B=0.2$~cm$^{-1}$. (c)-(f) Populations of rotational states with quantum number $j$ and magnetic quantum number $m$ for $N=1$ and $N=2$ at $t=200$\,fs. (g) Time-dependent population of molecular electronic excited states $P_{\mathrm{exc}}^{\mathrm{mol}}(t)$.}
    \label{fig:figure2}
\end{figure*}

%We want to study a sufficiently simple molecular model that is still able to capture basic mechanisms through which rovibronic and photonic DOFs interact in a gas-phase FP cavity.
For the simulations we consider a molecular Hamiltonian with interatomic distance $R_i$
and two electronic states, a ground and an excited electronic state with potential energy curves (PECs) $V_0$ and $V_1$ without intrinsic molecular nonadiabatic couplings.
To investigate the impact of vibrational motion, PECs are expanded around the Franck-Condon (FC) geometry $R_i^{(0)}$ up to second order in terms of nuclear displacements $Q_i = R_i-R_i^{(0)}$, $V_0(Q_i)=E_0+\frac{1}{2}\omega_v^2 Q_i^2$  and $V_1(Q_i)=E_1 + \kappa Q_i + \frac{1}{2}\omega_v^2 Q_i^2$,
%A nonzero potential energy gradient $\kappa$ at the FC point ($Q_i=0$) induces vibrational motion upon electronic excitation
%Accordingly, the molecules are described by a two-state vibronic Hamiltonian in the basis of electronic states,
\begin{align}
    \hat H^{(i)}_{\mathrm{mol}}  = \left[\hat T_{\mathrm{rot}}(\varphi_i,\theta_i) + \hat T_{n}(Q_i)\right]\mathbf{1}_{2\times2} + \begin{pmatrix}
V_0(Q_i) & 0 \\
0 & V_1(Q_i)
\end{pmatrix},
\end{align}
where $\mathbf{1}_{2\times2}$ denotes the two-dimensional identity matrix,
$\hat T_{n}(Q_i)$ is the vibrational kinetic energy operator
and $\hat{T}_{\mathrm{rot}}(\varphi_i,\theta_i) = \hat{L}^2_{\varphi_i\theta_i}/2\mathcal{I}$ is the rotational kinetic energy operator with moment of inertia $\mathcal{I}$ and rotational angular momentum $\hat{\vec L}_{\varphi_i\theta_i}$.
We assume constant moment of inertia $\mathcal{I}$ throughout this letter, effectively neglecting
centrifugal coupling between rotations and vibrations. We note that this coupling is very small compared to
other mechanisms, and that it could be easily reintroduced if needed. This approximation simplifies the analysis
of the energy transfer from the cavity and electronic excitations to the rotational degrees of freedom.

%According to Fig.~\ref{fig:figure_theo}(a) and the introduction, $\theta_i$ denotes the polar angle (out-of-plane) and $\phi_i$ the azimuthal angle (out-of-plane).
% I.e. we choose the two cavity polarization directions and mirror-to-mirror as the space-fixed frame and parameterize the molecular TDM vector within this frame in terms of spherical coordinates.
 To proceed, we perform an adiabatic separation between the ``fast'' electronic as well as cavity mode DOFs and the ``slow'' molecular vibrational and rotational DOFs \cite{fabri_bornoppenheimer_2021}. The representation of this adiabatic Hamiltonian, $\hat H_{\mathrm{ad}} = \hat H - \sum_{i=1}^N\left[\hat T_{\mathrm{rot}}(\varphi_i,\theta_i) + \hat T_{n}(Q_i)\right]$, in the basis of non-interacting cavity-ensemble states and in the single-excitation subspace, gives rise to a molecular Tavis-Cummings Hamiltonian \cite{tavis_exact_1968, vendrell_collective_2018}
\begin{eqnarray}
   \hat H_{\mathrm{MTC}}  = \begin{pmatrix}
\hbar\omega_c & 0  & \gamma_x^{(1)} & \gamma_x^{(2)} & \cdots \\
0 & \hbar\omega_c & \gamma_y^{(1)} & \gamma_y^{(2)} & \cdots \\
\gamma_x^{(1)} & \gamma_y^{(1)} & \Delta^{(1)} & 0 &   \cdots \\
\gamma_x^{(2)} & \gamma_y^{(2)} &  0 & \Delta^{(2)} &   \cdots \\ 
\vdots & \vdots & \vdots & \vdots & \ddots 
\end{pmatrix},
\end{eqnarray}
which has been extended to a two-dimensional FP cavity model. Its eigenvalues and associated eigenvectors correspond to the polaritonic (as well as dark) states and polaritonic potential energy surfaces (pPES), respectively \cite{feist_polaritonic_2018}. Here, $\Delta^{(i)}(Q_i)=V_1(Q_i)-V_0(Q_i)$ is the energy gap of the $i$-th molecule. The cavity photon energy $\hbar\omega_c$ is tuned resonant to the energy gap at the FC point ($Q_i=0$). $\gamma_x^{(i)}(\varphi_i,\theta_i)=g|\vec\mu_{01}^{(i)}(Q_i)|\cos\varphi_i\sin\theta_i$ $\gamma_y^{(i)}(\varphi_i,\theta_i)=g|\vec\mu_{01}^{(i)}(Q_i)|\sin\varphi_i\sin\theta_i$ are the orientation-dependent dipole couplings of the $i$-th molecule with the the $x$ and $y$-polarized cavity mode, respectively. We assume a constant TDM $|\vec\mu_{01}^{(i)}(Q_i)|=\mu_{01}$ and introduce $\lambda = g\mu_{01}$. The common assumption of fully aligned TDMs would result in an UP and LP branch which are separated by the collective Rabi splitting $\hbar\Omega_R = 2\lambda\sqrt{N}$ at $Q_1=\cdots=Q_N=0$.

The full analytical treatment of $\hat H_{\mathrm{MTC}}$ for arbitrary $N$ and including all rotational DOFs can be found in the Supplementary Material \cite{supp}.
% -> This is only the analytical consideration? 
Here, we limit the analytical considerations to the collective
effects emerging from molecular rotations inside the polarization plane by fixing the polar angles to $\theta_1=\cdots=\theta_N=\pi/2$ and the vibrational DOFs to the FC point.
For a single molecule rotating in the polarization plane, $\hat H_{\mathrm{MTC}}$ is rotationally invariant as any $\varphi_1$ dependence can be removed by a rotation around $\varphi_1$ in the basis of $x$- and $y$-polarized cavity modes: a single molecule feels the same cavity field strength for all possible orientations in the cavity polarization plane. Consequently, the single-molecule pPESs are independent of the in-plane rotation angle, 
$\varepsilon^{\pm}_{N=1} = \hbar\omega_c \pm \lambda$.\\
The situation drastically changes with just one more molecule inside the cavity. A strong dependence of the polaritonic energy landscape on the intermolecular alignment, $\cos^2(\varphi_1-\varphi_2)$, is found. Two pairs of upper polaritonic (UP) and lower polaritonic (LP) states are obtained analytically,
\begin{eqnarray}
    \varepsilon^{\pm,1}_{N=2} &=& \hbar\omega_c \pm \lambda \sqrt{1 - \sqrt{\cos^2(\varphi_1-\varphi_2)}},\label{eq:collCI1} \\ 
    \varepsilon^{\pm,2}_{N=2} &=& \hbar\omega_c \pm \lambda \sqrt{1 + \sqrt{\cos^2(\varphi_1-\varphi_2)}}. \label{eq:collCI2}
\end{eqnarray}
In this case, LP and UP states separated by the collective Rabi splitting $\hbar\Omega_R = 2\lambda\sqrt{2}$ are only found if the molecular TDMs are fully aligned, i.e. if $\cos^2(\varphi_1-\varphi_2)=1$. In the full rotational subspace this configuration corresponds to a line of degeneracy where the two remaining polaritonic states (with pPES given by $\varepsilon^{\pm,1}_{N=2}$) cross. The cooperative molecular rotation along $\varphi_1-\varphi_2$, which breaks the alignment, will lift this degeneracy [see Fig.~\ref{fig:figure_theo}(b), upper panel]. If vibrational DOFs are included, they can serve as a second branching coordinate, resulting in a conical intersection (CI) between a UP and LP surface \cite{supp}. 

This topology shares great similarity with LICIs and intrinsic molecular CIs including a divergent nonadiabatic coupling vector \cite{worth_beyond_2004} $\vec F_{ij} = \langle \psi_i|\vec \nabla_{\varphi} \psi_j\rangle$ among the adiabatic eigenstates $|\psi_j\rangle$ of $\hat{H}_{\mathrm{MTC}}$ near the intersection [see Fig.~\ref{fig:figure_theo}(b), lower panel].
We stress, however, that they are of \textit{collective} nature:
each molecule, individually coupled,
displays rotationally invariant pPESs along $\varphi_i$,
and the CIs locations depend on the relative in-plane orientations of the molecular ensemble.
Collective CIs (CCIs) have been previously reported for fully aligned molecular ensembles,
where collective nuclear displacements lead to \mbox{(pseudo-)}Jahn-Teller interactions among bright
and dark polaritonic states \cite{vendrell_collective_2018}.
We refer to the latter as vibrational-type CCI to distinguish them
from the rotational-type CCIs introduced in this letter.

Furthermore, for two molecules,
a second degeneracy is found at perpendicular molecular orientations such that each molecule interacts separately with the cavity through orthogonally polarized cavity modes. This results in two degenerate pairs of UP and LP states with the single-molecule Rabi splitting $\hbar\Omega_R = 2\lambda$.
In this way, all polaritonic bright and dark states are connected by CCIs involving collective rotations of the molecular ensemble. These results can be generalized to arbitrary numbers of molecules whose rotations are restricted to the polarization plane. For $N$ molecules, degeneracies between UP and LP states at fully aligned configurations as well as degeneracies inside the UP branch and LP branch exist, and are lifted along collective rotational coordinates. This enables the existence of CCIs which involve solely rotational DOFs for $N>2$. As an example, cuts through the pPESs of an ensemble of three molecules are shown in the Supplementary Material \cite{supp}.
In contrast, for vibrational CCIs, the points of intersection are reached through modulation of the diagonal entries
in $\hat{H}_{\mathrm{MTC}}$, whereas the off-diagonal couplings remain (almost) constant around the FC point~\cite{vendrell_collective_2018}.
This implies that the highest and the lowest energy pPES
do not cross, and the vibrational CCIs take place only within the dark-states manifold.

In the following, prototypical diatomic molecules with vibrational frequency $\omega_v=0.074$\,eV, tuning parameter $\kappa=0.1194$\,eV, an energy difference of 4.72\,eV between ground and excited electronic state at the FC point and a rotational constant of $B=0.2$\,cm${^{-1}}$ are considered. These model parameters reflect a typical UV excitation of a medium weight diatomic molecule, e.g. Na$_2$ ($B\approx0.15$\,cm${^{-1}}$) or Li$_2$ ($B\approx0.67$\,cm${^{-1}}$) \cite{irikura_experimental_2007}. The cavity photon energy is tuned resonant to the molecular electronic transition, $\hbar\omega_c=4.72$\,eV, with a Rabi splitting of $\hbar\Omega_R = 0.3$\,eV for fully aligned ensembles.
For better comparison of collective phenomena, the Rabi splitting is kept constant among different ensemble sizes by scaling the coupling strength by $1/\sqrt{N}$. Simulations without scaling of the coupling strength are shown in Fig.~S4 of the Supplementary Material \cite{supp}. A one-photon excitation of the cavity defines
the initial state in all numerical propagations, while keeping the molecules in their respective rovibronic ground states at $t=0$. This is akin to a short, impulsive excitation that creates a coherent
superposition of UP and LP states in the electronic and cavity subspaces.
%, involving all collective molecular excitations accessible
%from the isotropic ground state which can interact with the excited cavity mode
%through a nonzero total transition dipole moment. 

First, we follow the rotational response of the molecular ensemble by computing the ensemble total (in-plane + out-of-plane) kinetic rotational energy, $\langle \hat {T}_{\mathrm{\theta\varphi}}\rangle(t)=\langle\sum_i^N \hat{T}_{\mathrm{rot}}(\varphi_i,\theta_i)\rangle(t)$, as well as the kinetic rotational energy of the in-plane rotations only, $\langle\hat T_{\mathrm{\varphi}}\rangle(t)=\langle\sum_i^N \hat{T}_{\mathrm{rot}}(\varphi_i)\rangle(t)$, which is associated with the $z$-component of the molecular angular momentum operator. 

For a single molecule, the presence of a LICI between the UP and LP states strongly mixes rovibrational and electronic-photonic DOFs, which enables nonresonant energy transfer from the initial cavity excitation to the molecular out-of-plane rotation. Consequently, the total kinetic rotational energy in Fig.~\ref{fig:figure2}~(a) increases dramatically over the course of 200\,fs, populating a broad range of highly excited rotational states with quantum number $j$ [cf. Fig.~\ref{fig:figure2}~(c)]. This finding is in agreement with a previous study on diatomic molecules in classical laser light \cite{halasz_effect_2013}. Moreover, as collective CIs are absent, in-plane rotations cannot participate in the single-molecule case. In Fig.~\ref{fig:figure2}~(b), $\langle\hat T_{\varphi}\rangle(t)$ does not increase with time, only low-excited rotational states $m=0,\pm 1,\pm 2$ contribute [cf. Fig.~\ref{fig:figure2}~(d)]. %(\textcolor{red}{This follows from conservation of the $z$-component of the total photonic-molecular angular momentum, which has been worked out in the SI, necessary?}).
Accordingly, the common assumption of a FP cavity with one polarization direction is sufficient to describe quantum light-induced rotational dynamics of a single molecule in the cavity.

This approximation breaks down, however, when adding more molecules to the cavity, as can be seen from Fig.~\ref{fig:figure2}~(b). Now, \textit{collective} CIs, as introduced around equations \eqref{eq:collCI1} and \eqref{eq:collCI2}, are present along with \textit{individual} LICIs. This opens an additional channel for collective nonresonant rotational energy transfer to the molecular in-plane rotations. As a result, a fast increase of both, $\langle \hat {T}_{\mathrm{\theta\varphi}}\rangle(t)$ and $\langle \hat {T}_{\mathrm{\varphi}}\rangle(t)$, is observed for $N>1$. Not only rotational states with large $j$ but also with large $z$-component (up to $m=\pm 10$ for $N=2$) are populated, as shown in Fig.~\ref{fig:figure2}~(e)-(f). Inspecting the rotational dynamics serves as a proxy to assess the contributions of collective and individual CI dynamics involving in-plane, and out-of-plane rotations: The fraction $\langle \hat {T}_{\mathrm{\varphi}}\rangle(t)/\langle \hat {T}_{\mathrm{\theta\varphi}}\rangle(t)$ at 200\,fs increases with increasing number of molecule, reaching from 29\,\% ($N=2$) to 49\,\% ($N=10$). Thus, ultrafast dynamics along in-plane and out-of-plane rotational angles are virtually coexistent in molecular ensembles strongly coupled to two-dimensional FP cavity. In the picture of pPESs, as presented in Fig.~\ref{fig:figure_theo}(b), the initially prepared coherent polaritonic wavepacket will quickly dephase due to such collective rotational dynamics around CIs among polaritonic surfaces.

The present full-dimensional model of diatomic molecules in a FP cavity supports an additional source of nonadiabaticity, namely vibrational-type CCIs which contribute to the ultrafast nonradiative decay from the UP branch for $N>2$, and nonadiabatic nuclear dynamics inside the dark state manifold. In Ref.\,\cite{csehi_competition_2022}, the simultaneous presence of LICIs and vibrational-type CCIs has been shown to result in a cooperative mechanism in which relaxation through LICIs facilitates the participation of the dark state manifold. This argument can be extended to rotational-type CCIs which also connect the UP and the dark state manifold (cf. Fig.~\ref{fig:figure_theo} and Fig.~S1 in \cite{supp}). Such individual or collective nonradiative relaxation dynamics towards the dark state manifold mitigate the energy transfer towards molecular rotations, since this energy transfer channel is only open from bright polaritonic states. Furthermore, due to the increasing density of states, dark states can act as ``energy sinks" for larger molecular ensembles \cite{ribeiro_polariton_2018,fregoni_theoretical_2022}, ultimately slowing down the rotational energy transfer.

This effect is substantiated by a decrease in the rotational energy transfer rate in Fig.~\ref{fig:figure2}~(a)-(b) when increasing the number of molecules from $N=1$ to $N=10$. Both, the total and in-plane rotational dynamics are affected to a similar degree. Simultaneously, we observe in the collective regime that Rabi cycling dynamics (corresponding to oscillations with period $\tau_R\approx 13.8$\,fs, best visible for $N=1$) is quenched, and the overall population of bare molecular excited states $P_{\mathrm{exc}}^{\mathrm{mol}}(t)$ increases with increasing number of molecules in Fig.~\ref{fig:figure2}~(g). This indicates the enhanced decay to the dark state manifold.

\begin{figure}[h]
    \centering
    \includegraphics[width=0.45\textwidth]{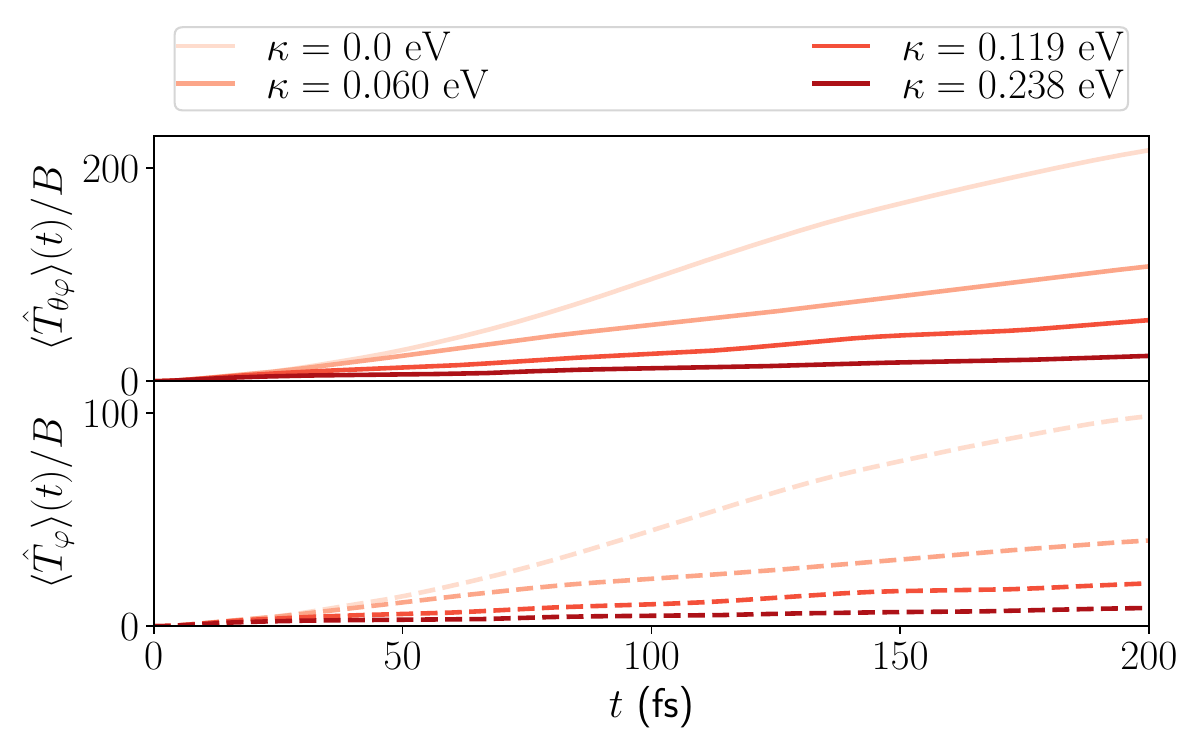}
    \caption{Total kinetic rotational energy $\langle \hat T_{\theta\varphi}\rangle(t)$ and in-plane rotational energy  $\langle \hat T_{\varphi}\rangle(t)$ for three molecules with varying gradient $\kappa$ of the molecular energy gap at the FC point.}
    \label{fig:figure3}
\end{figure}

Moreover, increasing the gradient of the molecular energy gap at the FC point (given by $\kappa$) for a fixed number of molecules increases the nonradiative relaxation rate through vibrational-type CCIs \cite{vendrell_collective_2018}. In agreement with the previous discussion, rotational energy transfer displays a high sensitivity to $\kappa$ in,
as seen in Fig.~\ref{fig:figure3}. Both, energy transfer to total and in-plane rotations, are slowed down significantly when the competing collective CI channels towards the dark states become more efficient.

In summary, diatomic gases enclosed in Fabry–Pérot (FP) cavities exhibit intricate coupling
among vibrational, rotational in-plane, and out-of-plane motion,
mediated by individual and collective cavity-matter interactions.
Consequently, examining the rotational dynamics offers a direct insight into the competition
among diverse cavity-enabled nonradiative relaxation pathways within the molecular ensemble upon cavity excitation.
Particularly, real-time wavepacket propagations unveil pronounced patterns in the distribution of molecular rotational levels and overall rotational energy transfer.
A notable observation is the substantial influence of the nonradiative relaxation rate to the dark states
on the efficiency of rotational energy transfer, 
providing a potential experimental proxy for probing these dynamics in pump-probe measurements.

\bibliography{cav_coll-rot}% Produces the bibliography via BibTeX.

\end{document}

\maketitle

\section{Analytical results}
%\subsection{Multi-mode Tavis-Cummings Hamiltonian}\label{ch:TC-Ham}
The molecular Tavis-Cummings Hamiltonian $\hat{H}_{\mathrm{MTC}}$ presented in the main text, takes the form of a $2 \times 2$ Hermitian block matrix, 
\begin{gather}
    \mathbf{H}_{\mathrm{MTC}}=\begin{pmatrix} \mathbf{\Pi} & \mathbf{G}\left(\lbrace \phi_i\rbrace;\lbrace \theta_i\rbrace\right)\\ \mathbf{G}^{\dagger}\left(\lbrace \phi_i\rbrace;\lbrace \theta_i\rbrace\right)& \mathbf{\Delta}\left(\lbrace Q_i\rbrace\right) 
    \end{pmatrix} \label{eq:SES-Ham-S0S1},
\end{gather}
with
\begin{gather}
    \mathbf{\Pi} = \begin{pmatrix} \hbar\omega_C & 0 \\ 0 & \hbar\omega_C\end{pmatrix}, \\
    \mathbf{G}\left(\lbrace \phi_m\rbrace;\lbrace \theta_m\rbrace\right) =  \lambda\begin{pmatrix} \cos\varphi_1\sin\theta_1 & \cos\varphi_2\sin\theta_2 & \dots & \cos\varphi_N\sin\theta_N \\
    \sin\varphi_1\sin\theta_1 & \sin\varphi_2\sin\theta_2 & \dots & \sin\varphi_N\sin\theta_N
     \end{pmatrix},\\
    \mathbf{\Delta}\left(\lbrace Q_i\rbrace\right) = \mathrm{diag}\,\left(\Delta^{(1)}(Q_1),\Delta^{(2)}(Q_2),\dots,\Delta^{(N)}(Q_N)\right),
\end{gather}
following the definitions of the main text. Note that out-of-plane angles $\lbrace \theta_i\rbrace$ are also considered here. \\
Following the approach introduced in Ref.\,\cite{wickenbrock_collective_2013}, a singular value decomposition (SVD) of the $2 \times N$ coupling matrix, $\mathbf{G}=\mathbf{U}\mathbf{\Sigma}\mathbf{W}^{\dagger}$ is performed. This facilitates the analytic treatment of the multi-mode Tavis-Cummings model. Here, $\mathbf{U}$ is a $2\times 2$ matrix, $\mathbf{W}$ a $N \times N$ matrix and $\mathbf{\Sigma}$ a $2 \times N$ diagonal rectangular diagonal matrix  with
\begin{gather}
\mathbf{\Sigma} = \mathrm{diag}(\lambda_1,\lambda_2,\dots, \lambda_{r})_{2\times N}.
\end{gather}
The singular values $\lambda_k$'s of $\mathbf{G}$ are obtained as the nonzero eigenvalues of $\sqrt{\mathbf{G}\mathbf{G}^{\dagger}}$ or $\sqrt{\mathbf{G}^{\dagger}\mathbf{G}}$. They are sorted such that $\lambda_1\leq\lambda_2\dots\leq\lambda_r$. Since $\mathbf{G}$ exclusively depends on angular variables, the singular values as well as the left- and right-singular vectors are functions of the $\phi$ and $\theta$ angles, i.e. $\lambda_k(\lbrace \phi_m\rbrace,\lbrace \theta_m\rbrace)$, $\mathbf{W}(\lbrace \phi_m\rbrace,\lbrace \theta_m\rbrace)$, and $\mathbf{U}(\lbrace \phi_m\rbrace,\lbrace \theta_m\rbrace)$. For simplicity of notation this dependence is dropped whenever obvious.  
The SVD allows for the definition of a basis of collective photonic excitations $\lbrace |\pi_k\rangle \rbrace$ and collective molecular excitations $\lbrace |\mu_k\rangle \rbrace$,
\begin{gather}
|\pi_k\rangle = \sum_{\alpha\in\lbrace x,y\rbrace}U_{\alpha k}|0;\alpha\rangle  \label{eq:coll-plm}  \\
    |\mu_k\rangle = \sum_{m=1}^N W_{mk}|m;0\rangle, \label{eq:coll-mol} 
\end{gather}
in which $\mathbf{G}$ becomes the rectangular diagonal matrix $\mathbf{\Sigma}$. The single excitation subspace basis states are denoted by $|0;\alpha\rangle$ and $|m;0\rangle$ where the single excitation is either located on the $\alpha$-polarized FP cavity mode or on the $m$-th molecule, respectively, while the remaining parts of the system are in their respective ground state. \\Transforming the multi-mode Tavis-Cummings Hamiltonian $\mathbf{H}_{\mathrm{MTC}}$ (\ref{eq:SES-Ham-S0S1}) into this basis requires transformation of the matrices $\mathbf{\Delta}$ and $\mathbf{\Pi}$ according to
\begin{gather}
  \bm{\Tilde \Delta} = \mathbf{W}^{\dagger}\bm{\Delta}\mathbf{W}\\
\bm{\Tilde \Pi} = \mathbf{U}^{\dagger}\bm{\Pi}\mathbf{U}. 
\end{gather}
Focusing on the impact of rotational DOFs, molecular vibrational DOFs are fixed to be uniform, i.e. $Q_1=\dots=Q_N=0$. In this case, $\bm{\Delta}$ is a multiple of the unity matrix and thus $\bm{\Tilde \Delta} = \bm{\Delta}$. Similarly, $\bm{\Tilde \Pi} = \bm{\Pi}$ since the cavity modes are degenerate. Note that the rank of the coupling matrix determines the number of singular values and thus, the number of bright and dark states: for $\mathrm{rank}({\mathbf{G}})=r\leq \mathrm{min}(2,N)$, $n_b =2r$ bright states and $n_d=2+N-2r$ dark states exist. Hence, in the present FP cavity model with two degenerate orthogonally polarized cavity modes at most $n_b=4$ polaritonic bright states can exist. Consequently, in this collective basis $\mathbf{H}_{\mathrm{MTC}}$ can be partitioned into a block-diagonal bright state part and a $n_d$-dimensional diagonal dark state part, $\mathbf{H}_{\mathrm{MTC}}=\mathbf{H}_{\mathrm{bright}}\oplus\mathbf{H}_{\mathrm{dark}}$ with $\mathbf{H}_{\mathrm{bright}} = \mathbf{h}_1 \oplus \dots \oplus \mathbf{h}_r$ and  $\mathbf{H}_{\mathrm{bright}} = \Delta \mathbf{1}_{n_d\times n_d}$. At the FC point, $\Delta=0$ holds. Each block can be diagonalized individually to obtain $2r$ eigenvalues of $H_{\mathrm{bright}}$, which read for $\Delta=0$,
\begin{align}
    \varepsilon^{\pm,1}_{N} &= \hbar\omega_c \pm \lambda_1(\lbrace \varphi_i\rbrace, \lbrace \theta_i\rbrace) , \\ 
    \varepsilon^{\pm,2}_{N} &= \hbar\omega_c \pm \lambda_2(\lbrace \varphi_i\rbrace, \lbrace  \theta_i\rbrace) 
\end{align}
Eigenstates (polaritons) are given by
\begin{gather}
    |\pm^{(k)}\rangle = \frac{1}{\sqrt{2}}\left(|\pi_k\rangle \pm |\mu_k\rangle \right). \label{eq:ses-eigVecs}
\end{gather}
The singular values $\lambda_1$ and  $\lambda_2$ are obtained as 
\begin{align}
    \lambda_1(\lbrace \varphi_i\rbrace, \lbrace \theta_i\rbrace) &= \lambda\sqrt{\frac{1}{2}\sum_i^N \sin^2(\theta_i) - \frac{1}{2}\sqrt{\sum_{i,j}^N \sin^2(\theta_i)\sin^2(\theta_j)\cos(2\varphi_i-2\varphi_j)}}\\
    \lambda_2(\lbrace \varphi_i\rbrace, \lbrace \theta_i\rbrace) &= \lambda\sqrt{\frac{1}{2}\sum_i^N \sin^2(\theta_i) + \frac{1}{2}\sqrt{\sum_{i,j}^N \sin^2(\theta_i)\sin^2(\theta_j)\cos(2\varphi_i-2\varphi_j)}}.
\end{align}
and lead to rotation-dependent Rabi splittings between the polaritonic branches. When molecular rotations are restricted to the cavity polarization plane (i.e. when $\theta_1=\dots=\theta_N=\pi/2$) these expressions simplify, 
\begin{align}
    \lambda_1(\lbrace \varphi_i\rbrace) &= \lambda\sqrt{\frac{N}{2} - \frac{1}{2}\sqrt{N+2\sum_{i<j}^N \left(2\cos^2(\varphi_i-\varphi_j)-1\right)}}\\
    \lambda_2(\lbrace \varphi_i\rbrace) &= \lambda\sqrt{\frac{N}{2} + \frac{1}{2}\sqrt{N+2\sum_{i<j}^N \left(2\cos^2(\varphi_i-\varphi_j)-1\right)}}.
\end{align}
Nonadiabatic coupling vectors $\vec F_{ij}$ among polaritonic states are calculated numerically from the unitary transformation matrix $\mathbf{C}$, which diagonalizes $\mathbf{H}_{\mathrm{MTC}}$, 
\begin{gather}
    \vec F_{ij} = \sum_k C_{ik}\left(\vec\nabla C_{jk}\right)
\end{gather}
via finite differentiation.
% \begin{gather}
%     \lbrace|1,m_l;0\rangle, |0,m_l;\alpha\rangle\rbrace \\
%     m_l = -\infty,\dots,-1,0,1,\dots,\infty \\
%     \langle m_k|\mathbf{GG}^{\dagger}|m_l\rangle = g^2 \langle m_k|\cos^2(\varphi) + \sin^2(\varphi)|m_l\rangle = g^2\delta_{kl} \\
%     |\mu_k\rangle = |1,m_k;0\rangle \\
%     |\pi_k\rangle = \sum_{m_l} c_j^{x,k}|0,m_j;x\rangle + c_j^{y,k}|0,m_j;y\rangle \\
%     \mathbf{G}^{\dagger}\mathbf{G} = g^2\begin{pmatrix}
%         \cos^2(\varphi)  & \sin(\varphi)\cos(\varphi) \\ \sin(\varphi)\cos(\varphi) & \sin^2(\varphi) 
%     \end{pmatrix}\\
%     g^2 \sum_{j} \langle m_i|\cos^2(\varphi)|m_j\rangle  c_j^{x,l} + \langle m_i|\sin(\varphi)\cos(\varphi)|m_j\rangle  c_j^{y,l} = g^2 c_j^{x,l} \\
%     g^2 \sum_{j} \langle m_i|\sin^2(\varphi)|m_j\rangle  c_j^{x,l} + \langle m_i|\sin(\varphi)\cos(\varphi)|m_j\rangle  c_j^{y,l} = g^2 c_j^{y,l} \\
%     c_j^{x,k} = \langle m_j|\sin(\varphi)|m_k\rangle \\
%     c_j^{y,k} = \langle m_j|\cos(\varphi)|m_k\rangle 
% \end{gather}

\section{Additional Figures}

\begin{figure}[H]
    \centering
    \includegraphics[width=0.8\textwidth]{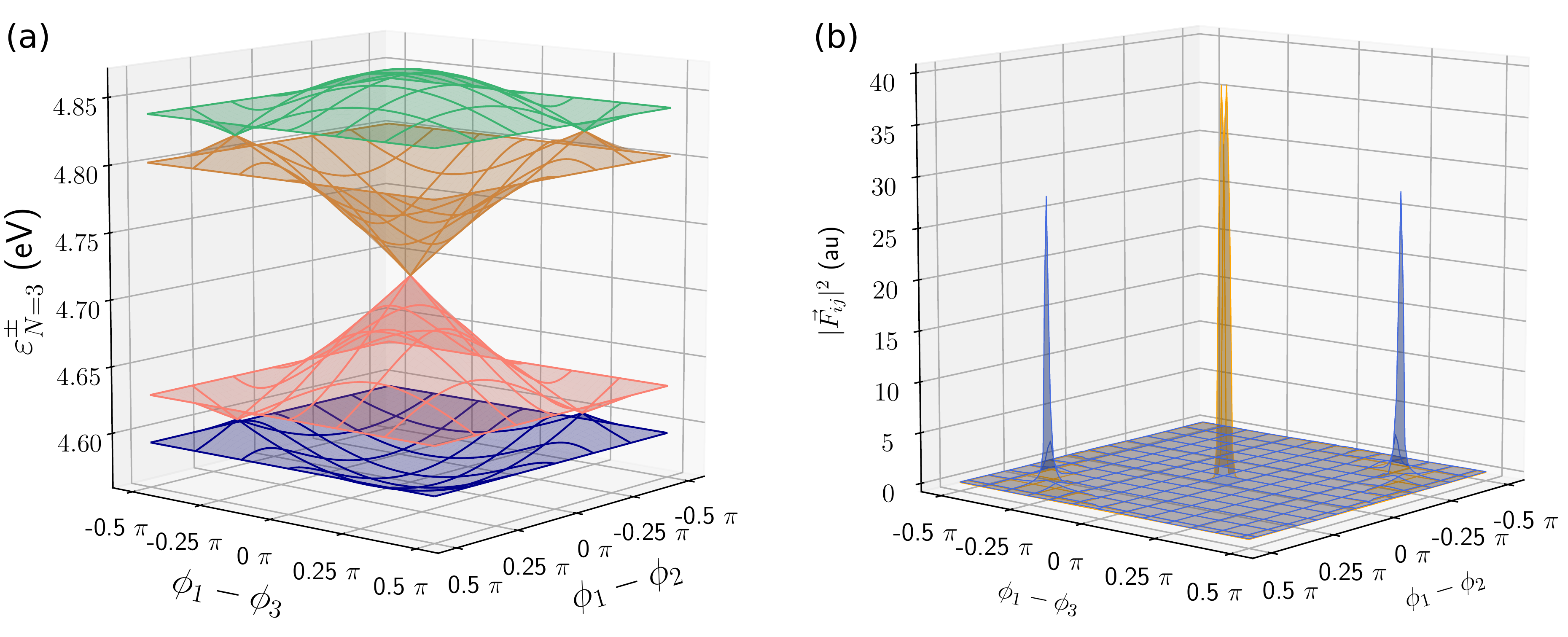}
    \caption{(a) pPESs for three molecules, $\varepsilon^{\pm}_{N=3}$, along relative angles $\varphi_1-\varphi_2$ and $\varphi_1-\varphi_3$. (b) Nonadiabatic couplings (NACs) $\vec F_{ij}$  among the two LP or UP states (blue), and between LP and UP states
    (orange) diverge at respective degeneracies of pPESs. pPESs and NACs are obtained from numerically diagonalizing $\hat H_{\mathrm{MTC}}$ (cf. main text) at the Franck-Condon point ($Q_1=Q_2=Q_3=0$).}
    \label{fig:enter-label}
\end{figure}
\begin{figure}[H]
    \centering
    \includegraphics[width=0.5\textwidth]{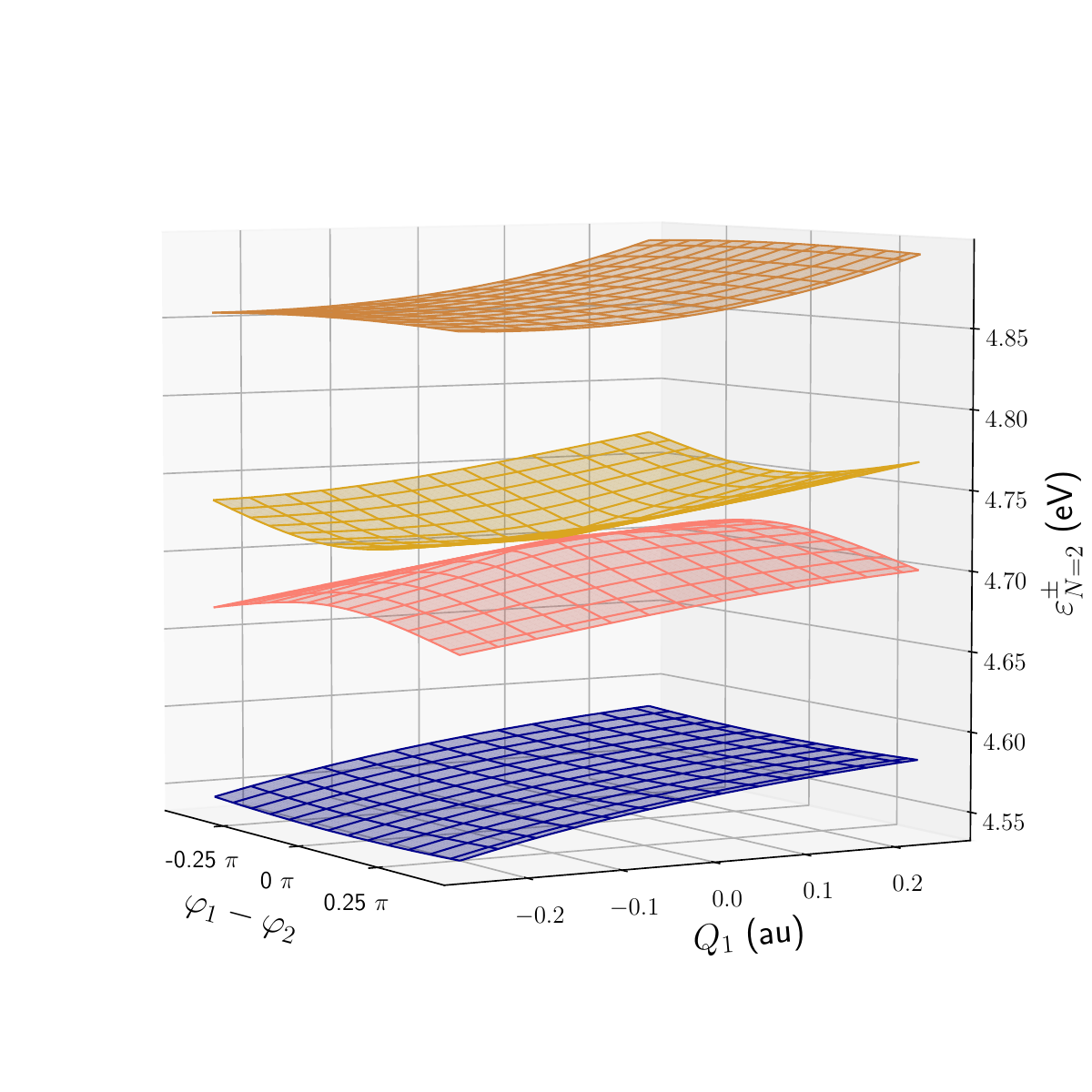}
    \caption{pPESs for two molecules, $\varepsilon^{\pm}_{N=2}$, as a function relative angles $\varphi_1-\varphi_2$ and $\varphi_1-\varphi_3$ and an intramolecular dimensionless vibrational coordinate $Q_1$.}
    \label{fig:enter-label}
\end{figure}

\begin{figure}[H]
    \centering
    \includegraphics[width=1.0\textwidth]{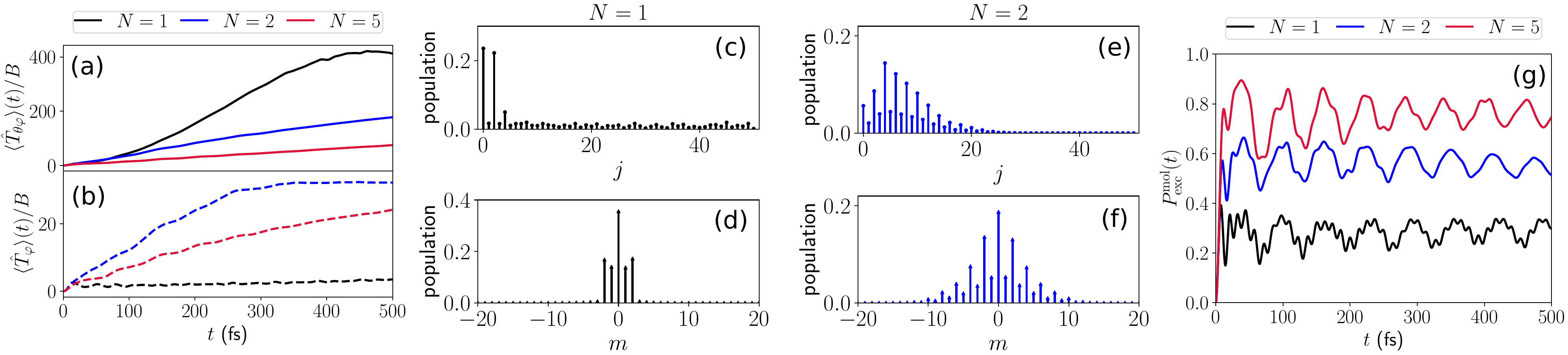}
    \caption{Molecular rotational dynamics after excitation of $x$-polarized cavity mode for various ensemble sizes for an extended propagation time of 500\,fs. (a)-(b) Total kinetic rotational energy $\langle \hat T_{\theta\varphi}\rangle(t)$ and in-plane rotational energy  $\langle \hat T_{\varphi}\rangle(t)$. Energies are given in units of the rotational constant $B=0.2$~cm$^{-1}$. (c)-(f) Populations of rotational states with quantum number $j$ and magnetic quantum number $m$ for $N=1$ and $N=2$ at $t=200$\,fs. (g) Time-dependent population of molecular electronic excited states $P_{\mathrm{exc}}^{\mathrm{mol}}(t)$.}
    \label{fig:enter-label}
\end{figure}

\begin{figure}[H]
    \centering
    \includegraphics[width=0.5\textwidth]{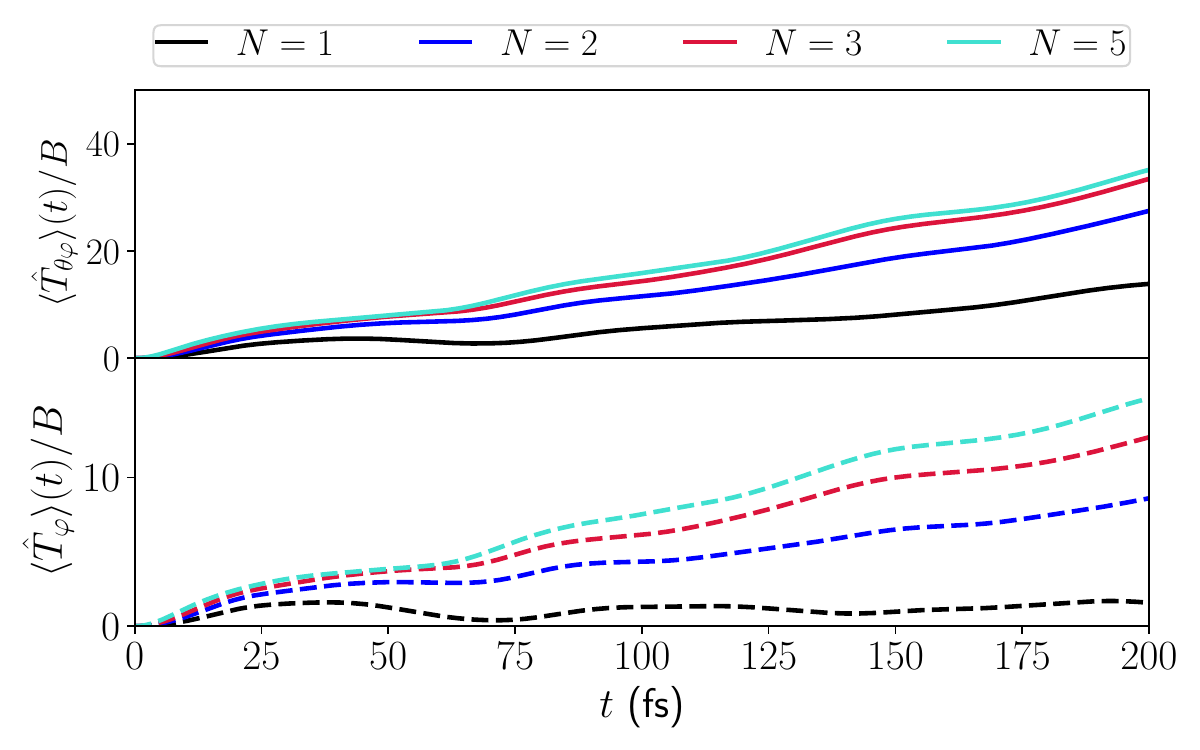}
    \caption{Impact of ensemble-size dependent collective Rabi splitting. Total kinetic rotational energy $\langle \hat T_{\theta\varphi}\rangle(t)$ and in-plane rotational energy  $\langle \hat T_{\varphi}\rangle(t)$ for $N=1,2,3,5$ and \textbf{without} $1/\sqrt{N}$-rescaling of coupling constant $g$. To avoid the ultrastrong coupling regime for larger $N$, a lower value of $g=0.067$\,eV is chosen. Energies are given in units of the rotational constant $B=0.2$~cm$^{-1}$.\\
    Compared to the scenario with constant Rabi splitting (cf. Fig.~2 in main text), rotational energy transfer rates continue to increase for $N>2$. For constant $g$ the energy gap between upper and lower polaritonic states increases with number of molecules. This leads to a suppression of nonradiative decay through vibrational CCIs for larger $N$ \cite{ulusoy_modifying_2019,csehi_competition_2022}, allowing collective rotational energy transfer to  increase beyond the onset of competing vibrational nonradiative relaxation channels.}
    \label{fig:enter-label}
\end{figure}

\begin{figure}[H]
    \centering
    \includegraphics[width=0.5\textwidth]{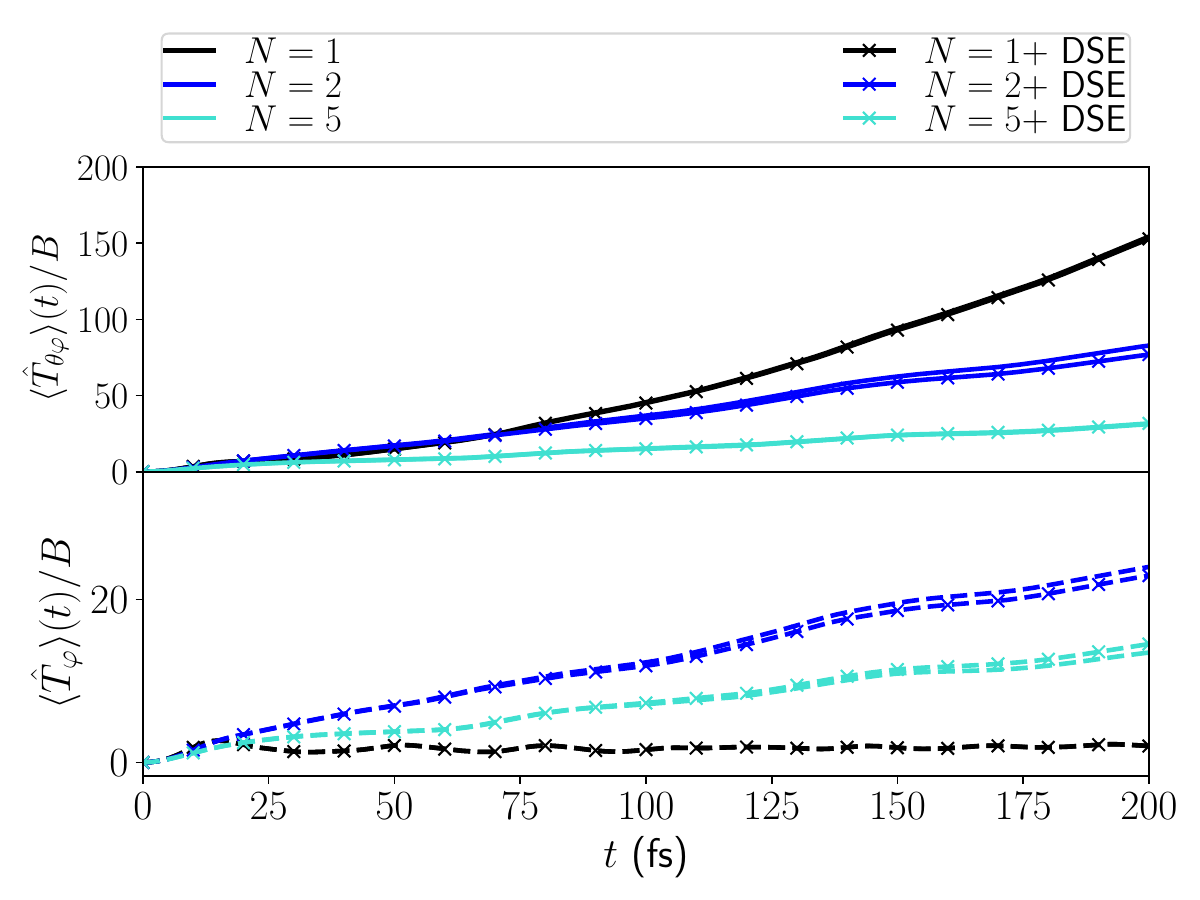}
    \caption{Impact of inclusion of dipole self energy term. Comparison between Fig.~2 (a)-(b) in main text with results obtained from propagations in which the dipole self energy (DSE) $g^2/\omega_c\sum_{\alpha=x,y}\left(\vec\epsilon_{\alpha} \sum_i^N\vec\mu^{(i)}\right)^2$ is included.}
    \label{fig:enter-label}
\end{figure}

% \begin{figure}[H]
%     \centering
%     \includegraphics[width=0.5\textwidth]{SI_Figures/Figure_pops.pdf}
%     \caption{Caption}
%     \label{fig:enter-label}
% \end{figure}

\bibliography{rots_SI}